\documentclass[12pt,tikz]{article}

\usepackage{tikz}
\usetikzlibrary{decorations.pathmorphing}

\usepackage{hyperref}

\usepackage{epsf}
\usepackage{amsmath,amssymb}
\usepackage{latexsym}
\usepackage{graphicx,color}
\usepackage{wrapfig}
\usepackage{latexsym}
\usepackage{graphics}

\usepackage{color}
\usepackage{delarray,color}
\usepackage{fancybox}  

\newcommand {\beq}{\begin{align}}
\newcommand {\eeq}{\end{align}}

\newcommand{\be}{\begin{equation}}
\newcommand{\ba}{\begin{align}}
\newcommand{\ea}{\end{align}}
\newcommand{\ee}{\end{equation}}

\newcommand{\ep}{\epsilon}

\newcommand{\beqa}{\begin{align}}
\newcommand{\eeqa}{\end{align}}

\newcommand{\unit}{\hbox to 3.8pt{\hskip1.3pt \vrule height 7.4pt
    width .4pt \hskip.7pt \vrule height 7.85pt width .4pt \kern-2.4pt
    \hrulefill \kern-3pt \raise 3.7pt\hbox{\char'40}}}

\def\matt[#1,#2,#3,#4]{\left(%
\begin{array}{cc} #1 & #2 \\ #3 & #4 \end{array} \right)}

\newcommand{\ket}[1]{{\left| #1 \right\rangle}}
\newcommand{\bra}[1]{{\left\langle#1\right|}}
\usepackage{tabularx}

\catcode`\@=11
\@addtoreset{equation}{section}

\catcode`@=12
\relax

\begin{document}

\begin{titlepage}

\setcounter{page}{0}

\renewcommand{\thefootnote}{\fnsymbol{footnote}}

\begin{flushright}
YITP-25-126
\end{flushright}

\vskip 1.35cm

\begin{center}
{\Large \bf 
Holography at Finite N: 

Breakdown of Bulk Reconstruction for Subregions
}

\vskip 1.2cm 

{\normalsize
Seiji~Terashima\footnote{terasima(at)yukawa.kyoto-u.ac.jp}
}

\vskip 0.8cm

{ \it
Center for Gravitational Physics and Quantum Information,

Yukawa Institute for Theoretical Physics, Kyoto University, Kyoto 606-8502, Japan
}

\end{center}

\vspace{12mm}

\centerline{{\bf Abstract}}

Within AdS/CFT, focusing on the AdS-Rindler wedge, we show that when $N$ is large but finite, correlation functions of reconstructed bulk operators grow exponentially with bulk momentum, overwhelming the usual $1/N$ suppression. 
The growth starts when the smeared operator's ultraviolet scale goes beyond a critical value
$\Lambda_{\mathrm{crit}} = \frac{2}{\pi}\ln N$, which is far below the Planck scale.
Above this logarithmic threshold, the large $N$ expansion ceases to be reliable, and the would-be bulk operators cannot be consistently defined as observables in the full quantum gravity theory. 
Since the AdS-Rindler wedge describes the near-horizon region of black holes, 
this result implies a sharp $\ln N$ cutoff for reconstructing bulk operators across horizons. 
This has a direct impact on whether and how information from the black hole interior is encoded—a central question in the black hole information paradox.

\end{titlepage}
\newpage

\tableofcontents
\vskip 1.2cm 

\setcounter{footnote}{0}

\section{Introduction and summary}
\label{sec:intro}

The AdS/CFT correspondence provides a non-perturbative definition of quantum gravity in spacetime with a negative cosmological constant~\cite{Maldacena:1997re} \cite{Gubser:1998bc} \cite{Witten:1998qj}.  
A central question is \emph{bulk reconstruction}:
How can the operators and states of the bulk gravity theory—viewed as a low energy effective description—be constructed from those of a well-defined (finite $N$ holographic) conformal field theory (CFT)?

In the free bulk theory, which corresponds to an $N=\infty$ CFT, the answer is known \cite{Banks:1998dd} \cite{Balasubramanian:1998sn} \cite{Bena:1999jv} \cite{Duetsch:2002hc} and explicitly 
supplied by the HKLL procedure~\cite{Hamilton:2006az} \cite{Kabat:2011rz}, which reconstructs free bulk fields by smearing CFT operators over an appropriate boundary domain.
Here, $N=\infty$ CFT means that we only consider the two-point functions, forgetting higher-order point functions by the large $N$ factorization.
Proposals that incorporate the $1/N$ corrections to this method have also been considered in \cite{Kabat:2011rz} \cite{Kabat:2013wga}\cite{Kabat:2015swa}.

Another important subject in the AdS/CFT correspondence is the interplay between \emph{subregions} on the boundary and in the bulk.  
Subregions are central to the study of the quantum–information–theoretic side of holography. 
In particular, a black hole horizon can be regarded as the boundary of a distinguished bulk subregion: near the horizon, the exterior geometry in Schwarzschild coordinates is well approximated by Rindler space, just as the causal wedge of a half–space in Minkowski space is described by Rindler coordinates.  
Consequently, a refined bulk reconstruction program that works within a given subregion is indispensable.  
Such a program is expected to shed light on the black hole information paradox and the structure of the black hole interior.  
The basic question can be phrased succinctly:
Which bulk operators can be reconstructed solely from CFT operators supported on the boundary region~$A$?

The entanglement wedge reconstruction (EWR) which
asserts that any operator supported inside the entanglement wedge, which is defined through the Ryu-Takayanagi surface \cite{Ryu:2006bv}, of a boundary region can be represented on the CFT operators supported on the corresponding subregion.
This was shown in the $N=\infty$ theory by~\cite{Dong:2016eik}, building on~\cite{Jafferis:2015del}, and is widely believed to remain valid, including 
$1/N$ corrections in the
$1/N$ expansion.\footnote{
This $N=\infty$ theory can be regarded as the generalized free theory, although it is not quantum field theory in $ d$ dimensions. 
}
A simple example of the entanglement wedge is the AdS-Rindler wedge in the global AdS space, which is the focus of this paper. We aim to discuss the fundamental aspects of bulk reconstruction for the subregion.
For this case, the EWR for $N=\infty$ theory is explicitly given by the HKLL bulk reconstruction \cite{Hamilton:2006az}.

Here, understanding how bulk reconstruction for a subregion behaves beyond the strict \( N = \infty \) limit becomes crucial for our comprehension of quantum gravity. In the case of entanglement wedge reconstruction (EWR), it has already been shown that the claims of~\cite{Dong:2016eik} must be modified once \( 1/N \) corrections are taken into account. Moreover, there are indications that the \( 1/N \) expansion itself may break down~\cite{Terashima:2020uqu, Terashima:2021klf, Sugishita:2022ldv, Sugishita:2023wjm, Sugishita:2024lee}. \footnote{These works have their origin in the operator formalism approach to understanding AdS/CFT \cite{Terashima:2017gmc, Terashima:2019wed}.} 
This suggests that the applicability of the bulk gravity theory may be fundamentally limited when subregions—or horizons—are involved.\footnote{
Previous works have shown that bulk reconstruction in AdS-Rindler or black hole backgrounds is considerably more subtle than in empty AdS. In particular, modes with exponentially small boundary imprint can obstruct the standard smearing-function construction, and the reconstruction of sub-AdS bulk locality from boundary local operators may require exponentially precise boundary data. Moreover, the large-spacelike-momentum behavior of thermal correlators has been studied in general relativistic quantum field theories and in holographic theories \cite{Leichenauer:2013kaa, Rey:2014dpa, Banerjee:2019kjh}.
}

In this paper, we show that operators constructed in the bulk gravitational theory on the AdS-Rindler wedge cannot be regarded as genuine bulk gravitational operators when realized in the true quantum gravity theory, namely a large but finite \( N \) holographic CFT.\footnote{In this work, we focus on the AdS–Rindler wedge in 
$d=2$ (for clarity), interacting bulk theories at large but finite $N$, and operator orderings where the reconstructed operator sits between two boundary insertions, which is precisely where the exponential enhancement becomes manifest.
}
More precisely, we show that correlation functions
of reconstructed bulk operators in CFT
grow exponentially with bulk momentum and overwhelm the standard $1/N$ suppression once the momentum exceeds a critical scale
\begin{equation}
  \Lambda_{\mathrm{crit}} \;=\; \alpha \,\ln N ,
  \qquad 
  \alpha = \tfrac{2}{\pi} ,
  \label{eq:logNcutoff-intro}
\end{equation}
where we set the AdS radius to unity and take the Planck mass to satisfy $M_{\mathrm{pl}}\sim N^{\frac{2}{d-1}}$.  
Beyond $\Lambda_{\mathrm{crit}}$, the large $N$ expansion diverges, signaling a breakdown of perturbation theory and invalidity of a (naive) low energy bulk gravity theory. 
Note that we presented this for the three-point functions, which are fixed by conformal symmetry except for an overall constant; thus, the result is not based on any assumptions on the quantum gravity. 

It is important to recognize that there exists a fundamental gap between the \( N \to \infty \) limit and the case of finite but large \( N \), particularly in the presence of horizons or when considering subregions. In such cases, the \( N \to \infty \) limit does not smoothly approximate the \( N = \infty \) theory. This observation does \emph{not} imply that the bulk gravitational description is invalid in the full quantum gravity theory; rather, it points to a fundamental limitation in applying naive semiclassical reasoning to situations involving horizons.

This distinction may reflect an essential feature of quantum gravity in the presence of horizons and could be crucial for resolving the black hole information loss paradox. In particular, we argue that the standard Rindler HKLL construction—based on EWR—fails for interacting bulk theories at finite \( N \). Similarly, popular frameworks such as the holographic quantum error correction code and subregion duality, which are based on the \( N = \infty \) structure, do not extend to finite \( N \) without significant modification.

Here, by "finite \( N \)" we do not mean non-perturbative effects in the \( 1/N \) expansion, such as the \( e^{-\mathcal{O}(N^2)} \) contributions considered in~\cite{Akers:2019wxj}. Such non-perturbative effects are typically extremely small and might not even be captured within the low energy bulk gravity theory. Rather, "finite \( N \)" is used here in contrast to the \( N = \infty \) limit, and refers to effects that are significantly larger than non-perturbative corrections—even larger than perturbative \( 1/N \) corrections—and which pose a fundamental challenge to bulk reconstruction for subregions.

\section{AdS-Rindler Bulk Operators and the log $N$ UV-Cutoff}


This section establishes a quantitative obstruction to bulk‐operator reconstruction in the AdS–Rindler wedge and identifies the scale at which the large-$N$ expansion necessarily breaks down.  
Our analysis unfolds in three steps:
We will use the HKLL bulk construction for a free scalar $\phi$ on the AdS–Rindler patch.
Throughout, we set the AdS radius to unity and take the Planck mass to scale as $M_{\mathrm{pl}}^{\,d-1}\!\sim N^{2}$.
Then, we will see correlation functions for $\phi^R$, which is a bulk local operator smeared by a UV regulator $\Lambda$, grow as
        \[
          \langle O_1\,\phi^R \,O_2\rangle
          \;\propto\;
          N^{-1}\exp \Bigl(\frac{\pi}{2} \Lambda \Bigr).
        \]
Requiring this factor to remain $\mathcal{O}(1)$ implies the critical condition
        \begin{equation}
          \Lambda
          \;\lesssim\;
          \frac{2}{\pi}\ln N.
          \label{eq:logNcutoff}
        \end{equation}
Beyond the threshold \eqref{eq:logNcutoff}, $1/N$ corrections fail to suppress higher–loop contributions, signalling that this bulk operator is not well-defined in the low energy bulk gravity theory in the quantum gravity, even though it exists in the (perturbative) bulk gravity theory.
Section 2 thus provides the technical foundation for the broader claim of this paper: certain bulk operators simply do not exist in the quantum theory defined on AdS–Rindler, and this obstruction propagates to the global AdS/CFT correspondence.

\subsection{Problem of bulk reconstruction for AdS-Rindler }

\begin{figure}[h]
\centering
\begin{tikzpicture}

  \draw[thick] (0,2) arc[start angle=90, end angle=270, radius=2];

  \draw[very thick, color=red!80!black] (0,-2) arc[start angle=270, delta angle=180, radius=2];

  \draw[thick, dashed, color=red!80!black] (0,-2) -- (0,2);

  \node at (1,0) {$\phi$};

  \filldraw (0.7,-0.5) circle (2pt);

  \node[text=red!80!black] at (2.4,1.4) {$A$};

\end{tikzpicture}
\caption{Bulk operator in AdS-Rindler wedge $M_A$.}
\label{bulkfig}
\end{figure}
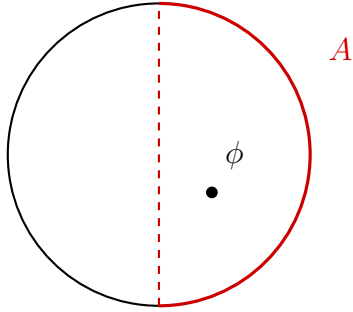

Let us consider the bulk operator $\phi$ which is supported on $M_A$,
where $M_A$ is the entanglement (causal) wedge of the subregion $A$ in CFT.
We take $M_A$ as the AdS-Rindler wedge. (Fig. \ref{bulkfig}.)
We will consider the 
reconstruction of the operator $\phi$ in CFT.
First, for the global AdS spacetime, we will have 
\( \phi_i^g({\cal O}) \), where \( {\cal O} \) is the CFT operators, by the HKLL bulk reconstruction formula \cite{Hamilton:2006az}.\footnote{
Precisely speaking, a bulk local operator is not a well-defined operator. The $\phi_i$ should be a smeared operator \cite{Morrison:2014jha}. 
}
The two-point functions of these are the same as the two-point function of the free bulk scalar theory on the AdS spacetime,
\begin{align}
    \bra{0} \phi_1 \phi_2 \ket{0} =\bra{0} \phi_1^g \phi_2^g \ket{0}, 
\end{align}
where $\ket{0}$ is the vacuum state of the bulk free theory or the CFT.
For the AdS-Rindler patch, \( M_A \), we will have 
\( \phi_i^R({\cal O}^R ) \), where \( {\cal O}^R \) is the CFT operator supported on the subregion $A$.\footnote{
More precisely, \( {\cal O}^R \) is the CFT operator supported on $D(A)$ where $D(A)$ is the domain of dependence of $A$. This is because operators in (non-trivial) CFT need smearing in the time direction \cite{Nagano:2021tbu}.  
We will use $A$ instead of $D(A)$ for notational simplicity.}
As before, we can show exactly
\begin{align}
    \bra{0} \phi_1 \phi_2 \ket{0} =\bra{0} \phi_1^R \phi_2^R \ket{0}.
\end{align}
Note that the vacuum state becomes a canonical ensemble if we consider CFT on region $A$, i.e., the Rindler patch.
Thus, for $N=\infty$, the AdS-Rindler reconstruction works well.
However, as we will see below, it does not work for finite $N$.

\subsubsection*{Claim of this paper}

Let us take the bulk operator $\phi$ as a bulk local operator smeared by a UV regulator $\Lambda$, for example $\phi \sim \int d x \, e^{-\frac{\Lambda^2}{2} (x-x_0)^2}\phi(t=t_0,x) $.
Then, we can show that 
the corresponding reconstructed CFT operator $\phi^R$ in $A$
satisfies 
\begin{align}
    \bra{0} {\cal O}_1 \phi^R {\cal O}_2 \ket{0} \sim \frac{1}{N} e^{\beta \Lambda},
\end{align}
where ${\cal O}_1, {\cal O}_2$ are generic CFT operators and $\beta$ is a non-negative constant, as we later find $\beta=\pi/2$.
Let us consider the case where the UV cutoff \( \Lambda \) is given by either  
\begin{align}
\Lambda = \alpha \ln N \quad \text{or} \quad \Lambda = (\ln N)^2.
\end{align}
In this case, the three-point function behaves as  
\begin{align}
\bra{0} {\cal O}_1 \phi^R {\cal O}_2 \ket{0}  \sim N^{\alpha \beta-1} \quad \text{or} \quad N^{\beta \ln N-1},
\end{align}
depending on the choice of the cutoff.
The key point here is that, although \( \Lambda \) is much smaller than the Planck mass,  
\(
M_{pl} \sim N^{2/(d-1)},
\)
the three-point function is still much larger than any fixed power of \( N \) for the $\Lambda = (\ln N)^2$ case, i.e.,  
\begin{align}
\bra{0} {\cal O}_1 \phi^R {\cal O}_2 \ket{0}  \gg N^{\gamma},
\label{ga}
\end{align}
for any constant $\gamma$.
For $\Lambda = \alpha \ln N$ case,
\eqref{ga} is satisfied for $\gamma < \alpha \beta-1$, which can be arbitrary large by taking $\alpha$ large.\footnote{
Note that such large values cause an overly large back reaction on the background, as the background is not reliable.
}
Note that $\beta=\pi/2$ is just a calculable numerical constant 
and $\alpha$ is an $\mathcal{O}(1)$ constant, we can choose as a parameter of the smearing.

In other words, although the three-point function is expected to be suppressed in the \( 1/N \) expansion, it not only fails to exhibit such suppression, but also results in an expression that cannot be interpreted as a well-defined operator in the low energy bulk theory.
This demonstrates that the CFT operator behaves correctly as the bulk operator one wishes to reproduce in the limit \(N = \infty\) (free bulk theory), but once interactions are introduced, it can no longer even be considered an operator in the bulk theory. 

Furthermore, in the AdS-Rindler bulk reconstruction, since this is the only operator that gives the two-point function of the corresponding bulk operator, it implies that such an operator does not exist in the well-defined version of the bulk theory of the AdS-Rindler patch.\footnote{
A perturbative bulk gravity theory will allow such an operator, but it will not be well-defined.
}
Here, we assume that the bulk theory of the AdS-Rindler patch 
will eventually be defined by the 
finite $N$ CFT on the region $A$
by the AdS/CFT correspondence \cite{Sugishita:2023wjm}.
We also note that the bulk theory of the AdS-Rindler patch considers a system at finite temperature $T=1/(2 \pi)$ \cite{Unruh:1976db}, which corresponds to the global vacuum. 

Thus, when considering the bulk theory as the low energy effective theory of the CFT on region $A$, we must take the UV cutoff $\Lambda$ of the theory to be $\Lambda \sim \ln N$, which is 
much smaller than the Planck scale.
(Alternatively, one may consider only operators without including horizon-to-horizon modes, but in that case, the bulk locality appears to be violated, even in an approximate sense. Other possibilities may exist, but we will not consider them here.)

\subsection*{Sketch of the proof}

Let us consider the AdS-Rindler patch, for which the metric is
\begin{align}
\label{Rindler-metric}
    ds^2=-\xi^2 dt_R^2+\frac{d\xi^2}{1+\xi^2}+(1+\xi^2) dH_{d-1}^2,
\end{align}
where $dH_{d-1}^2= d\chi^2 +\sinh^2\chi d\Omega_{d-2}^2$ is the metric of $(d-1)$-dimensional hyperbolic space $\mathbf{H}^{d-1}$
and $-\infty<t_R<\infty$, $0\leq\xi<\infty$, $0\leq\chi<\infty$.
In this paper, we set the AdS radius to unity.
The metric \eqref{Rindler-metric} is for $d>2$ and for $d=2$, the metric is 
\begin{align}
\label{Rindler-metric2}
    ds^2=-\xi^2 dt_R^2+\frac{d\xi^2}{1+\xi^2}+(1+\xi^2) d \chi^2,
\end{align}
where
$-\infty<\chi<\infty$.
The AdS-Rindler horizon is at $\xi=0$.
The map between this and the global coordinate, in which the metric becomes
    $ds^2=\frac{1}{\cos^2\! \rho}\left(-d\tau^2+d \rho^2+ \sin^2 \!\rho\, d\theta^2\right)$,
is given by
\begin{align}
\label{global-Rindler}
\begin{split}
\tan \tau &=\frac{\xi\sinh t_R}{\sqrt{1+\xi^2}\cosh \chi}, \quad 
    \cos\rho=\frac{1}{\sqrt{(1+\xi^2)\cosh^2 \chi+\xi^2 \sinh^2 t_R}},
    \\
    \tan \theta &=\frac{\sqrt{1+\xi^2}\sinh \chi}{\xi\cosh t_R}.
\end{split}
\end{align}

The bulk reconstruction for this AdS-Rindler patch is given in \cite{Hamilton:2006az}, and let us concentrate on the $d=2$ case for simplicity.
For the bulk reconstruction of a free scalar field $\phi$ with mass $m$, we first solve the equations of motion $(\Box-m^2) \phi=0$ and the positive frequency modes are given by
\begin{align}
    v_{\omega,\lambda}=e^{-i \omega t_R + i \lambda \chi} \tilde{\psi}_{\omega, \lambda}(\xi),
\end{align}
where $\omega$ is a positive continuous parameter and $\lambda \in \Re$. 
The wave function for the radial coordinate
$\tilde{\psi}_{\omega, \lambda}(\xi)$ is given by
\begin{align}
\label{psiR}
\tilde{\psi}_{\omega, \lambda}(\xi) 
=\frac{N_{\omega,\lambda}}{\Gamma(\nu+1)} \xi^{i\omega} (1+\xi^2)^{-\frac{i\omega}{2}-\frac{\Delta}{2}} 
    ~_2F_1\left(\frac{i\omega-i\lambda+\nu+1}{2} ,\frac{i\omega+i\lambda+\nu+1}{2}
  ;\nu+1;\frac{1}{1+\xi ^2}\right),
\end{align}
where 
\begin{align}
    \Delta:=1+\sqrt{m^2+1}, \qquad 
    \nu:=\Delta-1=\sqrt{m^2+1},
\end{align}
and the normalization constant $N_{\omega,\lambda}$ is fixed \cite{Sugishita:2022ldv} as 
\begin{align}
    N_{\omega,\lambda}=\frac{|\Gamma\left(\frac{i\omega-i\lambda+\nu+1}{2}\right)|\, |\Gamma\left(\frac{i\omega+i\lambda+\nu+1}{2}\right)|}{\sqrt{4\pi\omega} |\Gamma(i\omega)|}.
\label{nc}
\end{align}
The scalar field $\phi$ in the AdS-Rindler patch is expanded as
\begin{align}
\label{d=2_phi}
    \phi(t_R, \xi,\chi)=\int^{\infty}_{0}d\omega \int^{\infty}_{-\infty} d \lambda \frac{1}{\sqrt{2\pi}}
    \tilde{\psi}_{\omega, \lambda}(\xi) 
    \left[
    a_{\omega,\lambda} e^{-i\omega t_R +i \lambda \chi}+a^{\dagger}_{\omega,\lambda} e^{i\omega t_R -i \lambda \chi}
    \right].
\end{align}
where the operator $a_{\omega,\lambda}$ satisfies
\begin{align}
    [a_{\omega,\lambda}, a^{\dagger}_{\omega',\lambda' }]
    =\delta(\omega-\omega')\delta(\lambda-\lambda').
\end{align}
Using the BDHM map \cite{Banks:1998dd},
\begin{align}
    \lim_{\xi \to \infty} \xi^\Delta \phi(t_R, \xi, \chi)  =O_\Delta(t_R,\chi).
    \label{BDHM}
\end{align}
the corresponding CFT primary operator $O_\Delta(t_R,\chi)$ in the $N=\infty$ approximation can be written as
\begin{align}
    O_\Delta(t_R,\chi)= \frac{1}{\sqrt{2\pi}\Gamma(\nu+1)} \int^{\infty}_{0}d\omega \int^{\infty}_{-\infty} d \lambda \,
    N_{\omega,\lambda}
    \left[
    a_{\omega,\lambda} e^{-i\omega t_R +i \lambda \chi}+a^{\dagger}_{\omega,\lambda} e^{i\omega t_R -i \lambda \chi}
    \right],
\label{o1}
\end{align}
then, the bulk ladder operators $a_{\omega,\lambda}$ can be expressed by $O_\Delta$ as
\begin{align}
\label{a_by_O}
    a_{\omega,\lambda}= \frac{\sqrt{2\pi}\Gamma(\nu+1)}{N_{\omega,\lambda}} \int^{\infty}_{-\infty}\frac{d t_R}{2\pi} \int^{\infty}_{-\infty} \frac{d \chi}{2\pi} e^{i\omega t_R -i \lambda \chi} O_\Delta(t_R,\chi).
\end{align}

Let us consider the following bulk field:\footnote{
This operator involves fields located very close to the Rindler horizon, so there are certain subtleties associated with it. For this reason, we will slightly modify the operator in a more precise treatment later. However, for now, we proceed with the current form in order to illustrate the essential idea.
}
\begin{align}
\label{phil0}
   \phi_l= \int^{\infty}_{-\infty} d t_R \int^{\infty}_{-\infty} d \chi  e^{i\omega t_R -i \lambda \chi} \phi(t_R,\xi_0,\chi).
\end{align}
Here, we take 
$\xi_0$ is not too close to the asymptotic boundary,
then $\tilde{\psi}_{\omega,\lambda}(\xi_0)$ is not small and then
\begin{align}
\label{a_by_O2}
   \phi_l \sim
\tilde{\psi}_{\omega,\lambda}(\xi_0) a_{\omega,\lambda}\sim  a_{\omega,\lambda},
\end{align}
up to a numerical constant.
Using the AdS-Rindler bulk reconstruction, we can express it as a CFT operator for $N=\infty$: 
\begin{align}
\phi_l^R \sim
   \frac{1}{N_{\omega,\lambda}} 
   {\cal O}_{\omega, \lambda}, \,\,\,
   {\cal O}_{\omega, \lambda} \equiv \int^{\infty}_{-\infty} d t_R \int^{\infty}_{-\infty} d \chi  e^{i\omega t_R -i \lambda \chi} {\cal O}(t_R,\chi).
   \label{philtemp}
\end{align}
Now, we will take $\omega,\lambda$ such that
$|\lambda|-\omega >0 $ and $\omega,\lambda$ are very large,
which corresponds to the horizon-horizon mode \cite{Bousso:2012mh} (or the "tachyonic" mode) as explained in \cite{Sugishita:2022ldv}.
We also require that $|\lambda|-\omega $ is very large.
Because the normalization factor $N_{\omega,\lambda}$ behaves as
\begin{align}
N_{\omega,\lambda}
\rightarrow
\begin{cases}
\left( \frac{\omega^2-\lambda^2}{4} \right)^{\frac{\nu}{2}}
&\text{for}\quad
\omega^2 \geq \lambda^2\\
\left( \frac{\lambda^2-\omega^2}{4} \right)^{\frac{\nu}{2}} e^{-\frac{\pi}{2} (|\lambda|-\omega)}
&\text{for}\quad
\omega^2 < \lambda^2
\end{cases}
\label{N-NCFT}
\end{align}
in the limit $\omega, |\lambda| \rightarrow \infty$,
the reconstruction of $\phi_l$ contains the huge factor $1/N_{\omega,\lambda} \sim e^{\frac{\pi}{2} (|\lambda|-\omega)}$.
One might think that this factor would make the two-point function involving this operator large, but this is not the case, as bulk reconstruction is expected to work correctly for two-point functions.
Indeed, if we consider the bulk two point function $\langle 0| (\phi_l)^\dagger \phi_l |0 \rangle \sim
\langle 0| a_{\omega,\lambda}^\dagger a_{\omega,\lambda} |0 \rangle
$, which is regarded as the expectation value of the number operator for $a_{\omega,\lambda}$ in the thermal state in the Rindler patch, and then not large,
the corresponding expression in the CFT is 
\begin{align}
\langle 0| (\phi_l^R)^\dagger \phi_l^R |0 \rangle \sim
   \frac{1}{(N_{\omega,\lambda})^2} \langle 0| {\cal O}_{\omega, \lambda}^\dagger {\cal O}_{\omega,\lambda} |0\rangle,
\end{align}
and  $\langle 0| {\cal O}_{\omega, \lambda}^\dagger {\cal O}_{\omega,\lambda} |0\rangle \sim e^{-\pi (|\lambda|-\omega)}$ for $|\lambda|-\omega \gg 1 $ 
 as we will show later.
This implies that such ${\cal O}_{\omega,\lambda}$ has almost vanishing two-point functions.
This is also understood as follows: Let us consider $\delta_l \equiv \phi_l^R-\phi_l^G$ where $\phi_l^G$ is the global reconstructed CFT operator for the bulk operator $\phi_l$.
Then, $\delta_l \ket{0}= \bra{0} \delta_l=0$ and $\delta_l$ has vanishing two point functions \cite{Sugishita:2023wjm}, thus ${\cal O}_{\omega,\lambda} =N_{\omega,\lambda} (\delta_l + \phi_l^G)$ 
has almost vanishing two-point functions because 
${\cal O}_{\omega,\lambda} \ket{0}=N_{\omega,\lambda} \phi_l^G \ket{0}$. 
What is important here is that this CFT operator becomes very small when acting on the (global) vacuum, but there is no reason for it to be small when acting on states other than the vacuum. In fact, if we evaluate the three-point function $\langle 0| {\cal O}(t^1_R,\chi_1) \phi_l^R {\cal O}(t^2_R,\chi_2)|0 \rangle $ with generic points $t^i_R,\chi_i$,
we will have
\begin{align}
\langle 0| {\cal O}(t^1_R,\chi_1) \phi_l^R {\cal O}(t^2_R,\chi_2)|0 \rangle   \sim
   \frac{1}{N_{\omega,\lambda}} \langle 0| {\cal O}(t^1_R, \chi_1) {\cal O}_{\omega,\lambda} {\cal O}(t^2_R, \chi_2) |0\rangle  
\sim
   \frac{1}{N} e^{\frac{\pi}{2} (|\lambda|-\omega)},
\end{align}
which is suppressed by $1/N$, but enhanced by the factor $e^{\frac{\pi}{2} (|\lambda|-\omega)}$, as we will show later.\footnote{
The ordering of the operators is crucial.
}  

Thus, for $|\lambda|-\omega \simeq \alpha \ln N$,
we have $\langle 0| {\cal O}(t^1_R,\chi_1) \phi_l^R {\cal O}(t^2_R,\chi_2)|0 \rangle   \sim N^{\frac{\pi}{2} \alpha-1 }$, which invalidates the $1/N$ expansion for a sufficiently large $\alpha$.
Furthermore, the bulk local operator smeared with the UV cutoff $\Lambda$ will contain the mode $a_{\omega, \lambda}$ with $|\lambda|-\omega \leq \Lambda$, and then it is not a good operator in the $1/N$ expansion 
if $\Lambda \geq \frac{2}{\pi} \ln N$.

\subsection{Comments on the low energy gravity theory}

From the preceding arguments, it follows that the set of admissible operators and their properties can differ between the global AdS case and the AdS-Rindler case. Nevertheless, we still expect the AdS/CFT correspondence to hold in both settings. In particular, there should exist some effective low energy description of the (finite-temperature) CFT on the subregion \( A \), which is expected to be dual to a bulk gravity theory in AdS-Rindler space. This perspective was referred to as subregion complementarity in~\cite{Sugishita:2023wjm}.
In this context, it is expected that the UV cutoff of such a bulk theory must lie below the Planck scale, possibly of order \( \ln N \). As an alternative possibility, as discussed in~\cite{Sugishita:2022ldv}, one might consider excluding operators corresponding to horizon-to-horizon modes.\footnote{
The claim made in this paper for the AdS–Rindler wedge is expected to extend in a similar way to the description outside the horizon of an eternal black hole. However, the situation may be different for black holes formed through gravitational collapse. In particular, operators corresponding to horizon-to-horizon modes may not be includable at all in such cases.
} 
However, it remains unclear whether such a procedure can be consistently implemented.
In any case, it seems unlikely that the algebra of operators can be identified with that of gauge-invariant operators in the naive bulk gravity theory with a Planck-scale UV cutoff.

One implication of the claim that certain AdS-Rindler operators are absent in the full quantum gravity theory described by the holographic CFT concerns gravitational Wilson lines, which are a particular kind of gravitational dressing. These operators were explicitly constructed to low orders in perturbation theory in the Newton constant \( G_N \) in~\cite{Donnelly:2015hta, Donnelly:2016rvo}, and it is often assumed that such constructions can be extended more generally. However, what happens in full quantum gravity—especially when quantum effects are taken into account—is almost completely unknown. This issue is clearly non-trivial.
In fact, even in the seemingly simpler case of non-Abelian gauge theories, the expectation value of Wilson lines becomes very large in the confining phase, and diverges when the line is extended to infinity.
As we will argue below, our proposal indicates that while gravitational Wilson lines may exist under certain conditions, they may not be well-defined in general within the full quantum gravity theory, in a certain sense.

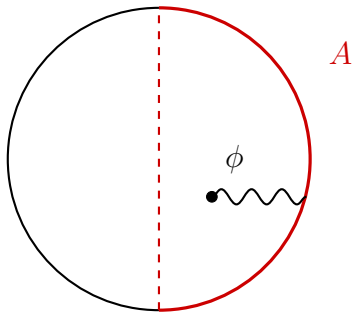
\begin{figure}[h]
\centering
\begin{tikzpicture}

  \draw[thick] (0,2) arc[start angle=90, end angle=270, radius=2];

  \draw[very thick, color=red!80!black] (0,-2) arc[start angle=270, delta angle=180, radius=2];

  \draw[thick, dashed, color=red!80!black] (0,-2) -- (0,2);

  \node at (1,0) {$\phi$};

  \filldraw (0.7,-0.5) circle (2pt);

  \draw[decorate, decoration={snake, amplitude=1mm, segment length=4mm}, thick]
        (0.7,-0.5) -- (1.95,-0.5);  

  \node[text=red!80!black] at (2.4,1.4) {$A$};

\end{tikzpicture}
\caption{Bulk smeared local field $\phi$ in the AdS-Rindler wedge, dressed by a gravitational Wilson line that ends on the boundary subregion $A$ on the $t_R=0$ slice. The line provides the diffeomorphism-invariant dressing, so that $\phi$ becomes gauge-invariant. The Rindler horizons bound the wedge (schematic).
}
\label{figp}
\end{figure}

Let us consider a bulk (smeared) local field $\phi$ with the gravitational Wilson line attached to it, such that it becomes a gauge invariant operator in the $t_R=0$ slice.
We assume that $\phi$ is in the AdS-Rindler patch and the gravitational Wilson line goes to the subregion $A$ in the asymptotic boundary, as in Fig. \ref{figp}.
Such an operator exists in the bulk gravity picture, provided the gravitational Wilson line is defined to all orders in $G_N$ and we regard this (semi-classical) gravity theory as a whole theory.
However, if the UV cutoff of the smearing of this bulk operator $\phi$ is large enough, the bulk operator is not well-behaved in the low energy bulk theory of the quantum gravity, because of the "divergent" behavior we have seen.
Thus, this gauge invariant, including the depicted gravitational Wilson line, is not described in the low energy bulk gravity theory due to the full quantum gravity effects.\footnote{
For some cases, the gravitational Wilson line will exist. For example, the wave packet type operator in global AdS, which is considered in \cite{Terashima:2020uqu, Terashima:2021klf, Terashima:2023mcr,  Tanahashi:2025fqi}, will be attached with two gravitational Wilson lines because the intersection of the shock wave caused by the wave packet and the $\tau=0$ slice is two lines. 
}

\begin{figure}[h]
\centering
\begin{tikzpicture}

\draw[thick] (0,0) circle(2);

  \node at (-1.2, 0.4) {$e^{i \phi}$};

  \filldraw (-1.4,0) circle (2pt);

  \draw[decorate, decoration={snake, amplitude=1mm, segment length=4mm}, thick]
        (-1.4,0) -- (2.0,0);
        
\draw[very thick, color=red!80!black] (2.0,-0.3) to [out=80,in=280] (2.0,0.3);

  \node[text=red!80!black] at (2.4,0) {$A$};

\end{tikzpicture}
\caption{Unitary operator $U = e^{i\phi}$ with the gravitational Wilson line ending on the very small boundary subregion $A$. This illustrates that, while such dressed operators can exist in the bulk description, for a large UV smearing scale $\Lambda$, they cease to be well-defined within the low energy effective theory.
}
\label{figu}
\end{figure}
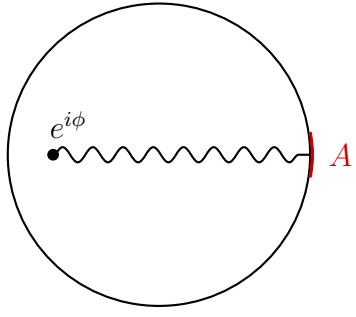

There is another argument indicating that the gravitational Wilson line may not be well-defined in some specific forms.
Let us consider the bulk smeared local operator $\phi$ with the gravitational line. 
Then, let us consider $e^{i\phi}$, where $\phi$ includes the gravitational Wilson line such that it becomes a unitary operator.
Here, the gravitational Wilson line ends on the very small boundary subregion $A$, as shown in Fig. \ref{figu}.
We expect that this unitary operator should be supported only in the subregion $A$ in the CFT. First, we note that $\bra{s} T_{00} \ket{s} = \bra{s} e^{-i\phi} T_{00}(x) e^{i\phi} \ket{s} $ if  $T_{00}(x)$ is in the subregion $\bar{A}$ and $\ket{s}$ is an arbitrary low energy state.
Here, $T_{00}(x)$ may be regarded as the insertion of $g_{00}$ very close to the boundary in the Fefferman-Graham gauge in the bulk gravity, which will commute with the bulk operators far from the boundary point $x$.
Because there will be no (dynamical) operator supported on a subregion which commutes with the energy momentum tensor at a point in the subregion, 
we find the bulk operator $e^{i \phi}$ will be reconstructed from the CFT operators supported on the very small subregion $A$ only.
However, this bulk operator is inserted at the region outside the entanglement wedge of $A$, even though it is reconstructed from the subregion $A$.
Thus, this contradicts the entanglement wedge reconstruction even at $N=\infty$, which means that we only consider the two-point function, if such an operator exists.
This suggests that such gravitational Wilson lines cannot exist.

\subsubsection*{On entanglement wedge reconstructions}

In the following, we discuss what the results of this paper imply about the validity of entanglement wedge reconstruction (EWR). First, as discussed in \cite{Terashima:2020uqu, Terashima:2021klf, Sugishita:2022ldv, Sugishita:2023wjm, Sugishita:2024lee}, the most standard version of EWR, as discussed in \cite{Dong:2016eik}, only works for the $N = \infty$ theory. In particular, it can be said that the properties assumed in the holographic error correction proposal of  \cite{Almheiri:2014lwa} \cite{Pastawski:2015qua} do not hold in our setup. 
In \cite{Sugishita:2022ldv}, the obstruction to EWR was identified as the so-called horizon-to-horizon modes \cite{Bousso:2012mh} in the bulk. These modes correspond precisely to the bulk gravitational modes on AdS-Rindler space that are shown in this paper to be non-existent as operator excitations.

On the other hand, subalgebra-based versions of EWR, such as those discussed in the context of algebraic entanglement entropy \cite{Harlow:2016vwg}, may still hold even when $1/N$ corrections are included, as discussed in \cite{Sugishita:2024lee} although these reconstructions do not exhibit the properties of holographic error correction \cite{Almheiri:2014lwa}, which is the base of the discussion in \cite{Harlow:2016vwg}.
For the subalgebra-based versions of EWR, the gravitational dressings, including the gravitational Wilson lines, are important.
While this paper focuses on the non-existence of certain bulk operators in AdS-Rindler gravity, this issue is not limited to EWR but reflects a more fundamental problem. This is because any operator constructed in bulk gravity on AdS-Rindler space should also exist in the global AdS bulk theory, and correspondingly, operators on the Rindler CFT should exist as operators in the global CFT.
For example, the bulk operator with a gravitational Wilson line considered in this subsection can be defined in global AdS. However, when attempting to reconstruct it from the global CFT, one finds that it cannot be interpreted as a bulk operator in the corresponding gravitational background. In other words, the CFT reconstruction fails to yield a bulk interpretation for such operators.

\section{Details}

In this section, we present a precise and detailed analysis building upon the discussion in the previous section.

Let us consider the following bulk (low energy) operator:

\begin{align}
\label{phil}
   \phi_l  (\omega_0, \lambda_0, t_R^0,\chi^0,a) &= \int^{\infty}_{-\infty} d t_R \int^{\infty}_{-\infty} d \chi  e^{-\frac{1}{2 a^2}(t_R-t_R^0)^2 -\frac{1}{2 a^2}(\chi-\chi^0)^2 +i\omega_0 t_R -i \lambda_0 \chi} \phi(t_R,\xi_0,\chi),
\end{align}
which 
is written using the creation and the annihilation operators as
\begin{align}
\label{phil2}
\phi_l(\omega_0,\lambda_0,t_R^0,\chi^0,a)
=&\,{a^2\over\pi}\sqrt{1\over2\pi}
\int_0^\infty d\omega \int_{-\infty}^{\infty} d\lambda\,
\widetilde\psi_{\omega,\lambda}(\xi_0)
\nonumber\\
&\times\Big[
 e^{-{a^2\over2}(\lambda-\lambda_0)^2-{a^2\over2}(\omega-\omega_0)^2
    -i\omega t_R^0+i\lambda\chi^0}
 a_{\omega,\lambda}
\nonumber\\
&\hspace{1.0cm}
+e^{-{a^2\over2}(\lambda+\lambda_0)^2-{a^2\over2}(\omega+\omega_0)^2
    +i\omega t_R^0-i\lambda\chi^0}
 a^\dagger_{\omega,\lambda}
\Big] .
\end{align}
for the free bulk theory.
Here, $a$ is the length scale of the smearing for the $t_R, \chi$ direction \cite{Terashima:2023mcr}.
We take the parameters such that 
\begin{align}
    |\lambda_0| / \omega_0 > 1, \,\,  \omega_0 \gg 1/a^2, \,\,\, a \ll 1,
\end{align}
(we will take $ |\lambda_0| \rightarrow \infty$ with $ |\lambda_0|/\omega_0 $ fixed finite),
and
\begin{align}
    \frac{1}{1+(\xi_0)^2} > \frac{\lambda^2-\omega^2}{\lambda^2}.
    \label{hhc}
\end{align}
The latter condition implies that there exists an (horizon-horizon) null-geodesic through the point $\{  t_R^0,\chi^0, \xi_0 \}$, so the $\xi_0$ is not close to the asymptotic boundary.
This bulk operator can be regarded as an operator creating wave packets at $\{ t_R^0,\chi^0, \xi_0 \}$ with the energy and the momentum $\{ \omega_0, \lambda_0 \}$, 
and roughly speaking, $\phi_l \sim e^{-i\omega_0 t^0_R +i \lambda_0 \chi^0} a_{\omega_0, \lambda_0}$ 
because, as we will see in section \ref{ab},  $\tilde{\psi}_{\omega, \lambda}(\xi_0)$ 
behaves as $|\lambda_0|^{-\frac{1}{2}}$.
Note that we assumed $\omega_0 \gg 1/a^2$, which is stronger than $\omega_0 a \gg 1$.
This condition will be used later.

We will examine correlation functions that include this operator, particularly their behavior in the large $\lambda$ limit.
In what follows, we focus on exponential factors in $|\lambda|$.  Factors that are only powers of the large momenta will be kept when they are useful, but they will not affect the exponential estimate.

First, we will check that two-point functions including $\phi_l$ in the free bulk theory do not have the exponential factor, although this statement is almost trivial.
We will consider, for example, $\langle 0| (\phi_l)^\dagger \phi_l |0 \rangle$, where $\ket{0}$ is the global vacuum.
Although $\phi_l$ in \eqref{phil} is expressed in the AdS-Rindler coordinate, we can rewrite it in the global coordinate. The map between them is smooth except for the horizon. Because $\xi$ is fixed to $\xi_0$ in  \eqref{phil}, only the singular points are $|t_R|=\infty$ or $|\chi|=\infty$, but these points are Gaussian suppressed.\footnote{
More precisely, we find that near these points the integrand is a smooth ($C^\infty$) function.
For example, near $\chi=\infty$, which implies $\theta \simeq \pi/2$, we see that $\tilde\theta \equiv \pi/2-\theta \sim e^{-\chi}$ and $\int d \chi e^{-(\chi-\chi_0)^2/(2 a^2)} \sim \int d \tilde\theta \,\tilde\theta^{-\ln \tilde\theta/(2 a^2)-1}$.
This integrand is not an analytic function, but a $C^{\infty}$ function. 
}
Thus, the two-point functions including $\phi_l$ in the free bulk theory do not have the exponential factor.
The higher point function will also behave regularly, but in order to define it, we need to include a gravitational dressing to $\phi_l$ for the gauge invariance, although it is not unique.
In the bulk (perturbative) perspective, a gravitational dressing with the correct $G_N (\simeq 1/N^2)$ expansion can be expected to exist.
However, it is not guaranteed in quantum gravity (i.e., a finite $N$ holographic CFT).

\subsection*{Reconstructed operator in CFT}

According to \cite{Hamilton:2006az, Morrison:2014jha}, we have the CFT operator $\phi_l^R$ in the $N=\infty$ approximation,
which is supported only in the region $A$, corresponding to the bulk operator $\phi_l$.
Indeed, $\phi_l^R$ is formally obtained by replacing $a_{\omega, \lambda}$ and $a^\dagger_{\omega, \lambda}$ in $\phi_l$ in \eqref{phil2}
by the CFT primary operator ${\cal O}$ using \eqref{a_by_O}.
Explicitly, we have
\begin{align}
    \label{philO}
   \phi_l^R = &
   \frac{a^2 \Gamma (\nu+1)}{4 \pi^3} \int^{\infty}_{0} 
   d \omega  \int^{\infty}_{-\infty} d \lambda  \, \frac{\tilde{\psi}_{\omega, \lambda}(\xi_0)}{N_{\omega,\lambda}} \, 
     \nonumber \\
   &  \times \left(  e^{-\frac{a^2}{2}(\lambda-\lambda_0)^2 -\frac{a^2}{2}(\omega-\omega_0)^2-i\omega t^0_R +i \lambda \chi^0}  \int^{\infty}_{-\infty} d t_R \int^{\infty}_{-\infty} d \chi e^{i\omega t_R -i \lambda \chi} O_\Delta(t_R,\chi)   + \cdots \right),
\end{align}
where $\cdots$ represents the similar term for $a^\dagger$.
Here, we will replace $\int_0^{\infty} d \omega$ in \eqref{philO} to $\int_{-\infty}^{\infty} d \omega$ because the difference is highly suppressed by the Gaussian factor $e^{-a^2(\omega_0)^2/2}$, which we will neglect.
We will also neglect the term that corresponds to $a^\dagger$ because it is suppressed by a factor  $e^{-a^2(\omega_0)^2/2}$.
More explicitly, this omitted term is not a small correction to the same wave packet.  It is the packet centered at the opposite point
$(\omega,\lambda)=(-\omega_0,-\lambda_0)$ in Fourier space.  Since the Rindler frequency label in the mode expansion satisfies
$\omega>0$, this packet has no saddle point in the integration region.  Its contribution comes only from the endpoint region near
$\omega=0$ and is bounded, up to factors which are powers of the large momenta and up to possible exponential factors already present in
$1/N_{\omega,\lambda}$, by
\begin{align}
   \exp\left[-{1\over 2}a^2\omega_0^2+O(\omega_0)\right].
\end{align}
Thus, under the assumption $a^2\omega_0\gg1$, it is exponentially smaller than the positive-frequency packet.  This is the sense in which the
$a^\dagger$ term is neglected below.

In Appendix \ref{ab}, it is shown that 
$\frac{\tilde{\psi}_{\omega_0, \lambda_0}(\xi_0)}{N_{\omega_0,\lambda_0}} 
    \sim e^{-i \omega_0 t_c+i \lambda_0 \chi_c} 
    \frac{1}{(|\lambda_0|-\omega_0)^{\nu+\frac{1}{2}}}$
for
$ |\lambda_0| \rightarrow \infty$ with $ |\lambda_0|/\omega_0 >1 $ fixed finite, as seen from \eqref{psitab} \eqref{psiN}.
Here, $t_c=t_r -i \frac{\pi}{2}, \chi_c=\chi_r -i \frac{\pi}{2} \operatorname{sgn} (\lambda_0)$ and $t_r, \chi_r$ are given implicitly in \eqref{psiN} as a function of $\omega_0/\lambda_0$.\footnote{
We assumed $\omega_0 \gg 1/a^2$ and 
$\lambda_0 \gg 1/a^2$, which implies the Gaussian factor in \eqref{philO} is not changed significantly.}
The most important part of $\frac{\tilde{\psi}_{\omega, \lambda}(\xi_0)}{N_{\omega,\lambda}}$ for the limit is given by $e^{-i \omega t_c+i \lambda \chi_c} 
    \frac{1}{(|\lambda_0|-\omega_0)^{\nu+\frac{1}{2}}}$.
Now we can expand some part of $\frac{\tilde{\psi}_{\omega, \lambda}(\xi_0)}{N_{\omega,\lambda}}$ in \eqref{philO}
around $(\omega,\lambda)=(\omega_0, \lambda_0)$
as 
\begin{align}
     e^{i \omega t_c-i \lambda \chi_c}  \frac{\tilde{\psi}_{\omega, \lambda}(\xi_0)}{N_{\omega,\lambda}} 
    = 
    C_{\xi_0, \omega_0, \lambda_0}+ \psi_{\xi_0, \omega_0, \lambda_0}^{(1,0)} (\omega-\omega_0)+ \psi_{\xi_0, \omega_0, \lambda_0}^{(0,1)} (\lambda-\lambda_0)+\cdots 
    , 
\end{align}
where $C_{\xi_0, \omega_0, \lambda_0} =\frac{\tilde{\psi}_{\omega_0, \lambda_0}(\xi_0)}{N_{\omega_0,\lambda_0}}  e^{i \omega_0 t_c-i \lambda_0 \chi_c} = {\cal O} \left(\frac{1}{(\omega_0)^{\nu+\frac{1}{2}}} \right) $ and $\psi_{\xi_0, \omega_0, \lambda_0}^{(n,m)}  ={\cal O} \left(\frac{1}{(\omega_0)^{\nu+\frac{1}{2} +n+m}} \right)$.
Then, we have 
\begin{align}
    \label{s1}
   \phi_l^R = & \frac{a^2 \Gamma (\nu+1)}{4 \pi^3} \int^{\infty}_{-\infty} d t_R \int^{\infty}_{-\infty} d \chi 
   \left( C_{\xi_0, \omega_0, \lambda_0}+ \frac{\psi_{\xi_0, \omega_0, \lambda_0}^{(1,0)}}{a^2} \frac{\partial}{\partial \omega_0}+\cdots 
   \right) \nonumber \\
   & \times \left( e^{i\omega_0 (t_R-t_R^0-t_c) -i \lambda_0 (\chi-\chi_0-\chi_c)-\frac{1}{2 a^2} ((t_R-t_R^0-t_c)^2+(\chi-\chi_0-\chi_c)^2) } O_\Delta(t_R,\chi)   
   \right),
\end{align}
where we have dropped the terms for $a^\dagger$, which are much smaller than the above terms. We will use this expression for $\phi_l^R$ below.
Note that there remains the large factor $e^{\frac{\pi}{2} (|\lambda_0|-\omega_0)}$ 
in \eqref{s1} as $e^{\omega_0 \operatorname{Im} (t_c) - \lambda_0 \operatorname{Im}(\chi_c)}$.

\subsection*{Two-point function in CFT}

The two-point function should be reproduced as
\begin{align}
    \langle 0| \phi_l \phi_1 |0 \rangle = \langle 0| \phi_l^R \phi_1^{CFT}  |0 \rangle, 
\end{align}
where $\ket{0}$ is the global vacuum of the bulk theory or the CFT, $\phi_1$ is a bulk operator, and $\phi_1^{CFT}$ is the corresponding CFT operator reconstructed by the global HKLL. 
If $\phi_1$ is supported on the AdS-Rindler patch, $\phi_1^{CFT}$ can be the corresponding CFT operator reconstructed by the AdS-Rindler HKLL.

Such a two-point function does not exhibit exponential growth in momentum, whereas the AdS-Rindler reconstructed CFT operator contains such a factor. In the following, we examine the reason why these seem to give different answers at first glance. To this end, we take ${\cal O}_{\Delta} (t_R=0,\chi=0)$ as $\phi_1^{CFT}$ and consider 
$\langle 0|  \phi_l^R {\cal O}_{\Delta}(0,0)  |0 \rangle$.
The two-point function for these fields is given by
\begin{align}
\langle 0|  {\cal O}_{\Delta}(t_R,\chi)  {\cal O}_{\Delta}(0,0)  |0 \rangle  =\frac{1}{(4 \sinh{\frac{u}{2}} \sinh{\frac{v}{2}} )^\Delta},
\end{align}
where 
\begin{align}
    u=t_R-\chi, v=t_R+\chi.
\end{align}
This can be derived by the coordinate transformation of the two-point function in the global coordinate (cylinder) or
the two-point function for the thermal state with temperature $1/(2 \pi)$ in the Rindler patch. 
The ordering of the operators is fixed by replacing $u,v$ with $ u-i\epsilon, v-i\epsilon$, although we will omit this $\epsilon$ for notational simplicity.
Then, we find 
\begin{align}
    \langle 0|  \phi_l^R {\cal O}_{\Delta}(0,0)  |0 \rangle = 
   & \frac{a^2 \Gamma (\nu+1)}{4 \pi^3} \left( C_{\xi_0, \omega_0, \lambda_0}+ \frac{\psi_{\xi_0, \omega_0, \lambda_0}^{(1,0)}}{a^2} \frac{\partial}{\partial \omega_0}+\cdots 
   \right) \nonumber \\
   &\times  \int^{\infty}_{-\infty} d u \int^{\infty}_{-\infty} d v \,
   \exp\left[i (u-u_0) p^u_0 + i (v-v_0) p^v_0
        -\frac{(u-u_0)^2+(v-v_0)^2}{4 a^2}\right] \nonumber \\
   &\times {1\over (4 \sinh{\frac{u}{2}} \sinh{\frac{v}{2}} )^\Delta} .
\end{align}
where $p^u_0= (\omega_0+\lambda_0)/2, p^v_0= (\omega_0-\lambda_0)/2$ and $u_0=t_R^0-\chi_0+t_c-\chi_c, v_0=t_R^0+\chi_0+t_c+\chi_c$.
Note that $p^u_0 p^v_0 < 0$ by the assumption we made.
To evaluate this, we will follow \cite{Terashima:2023mcr} \cite{Tanahashi:2025fqi}.
By shifting the integration contour for \(u\) to \(u \in \mathbb{R} + 2 i  a^2 p_0^u +i \operatorname{Im}(u_0) \) and rewriting $u=s+2 i  a^2 p_0^u+i \operatorname{Im}(u_0)$, the exponential factor in the integrand becomes \(  e^{-\frac{1}{4 a^2} (s-\operatorname{Re} (u_0)   )^2 -(a p_0^u)^2 +i u_0 p^u_0  }   \). 
Since the Gaussian factor \(e^{-(a p_0^u)^2} \) suppresses the contribution along this path,\footnote{
More precisely, 
$e^{-(a p_0^u)^2- \operatorname{Im}(u_0) p^u_0 }$ is small because of our assumption $\omega_0 a^2 \gg 1$.
} it can be neglected compared with the contributions from the poles of $\frac{1}{( \sinh{\frac{u}{2}} )^\Delta}$.\footnote{
If $\Delta$ is not an integer, those are branching points instead of the poles. As we have seen in \cite{Terashima:2023mcr}, the branching point cases can be treated as for the poles. 
}
Here, the direction of the deformation of the path is given by the sign of $p_0^u$ because $2 a^2 p_0^u +\operatorname{Im}(u_0) \simeq 2 a^2 p_0^u$.
Then, the dominant contribution is the one from the pole at \(u = \ i\epsilon \) if $p_0^u >0 $ or at \(u = \ -i(2 \pi -\epsilon) \) if $p_0^u  < 0 $. 
This is because contribution comes from the pole at \(u = \ \mathop{\mathrm{sgn}} (p_0^u) \, i(2 n \pi -\epsilon) \) with $n \in \mathbf{Z}_{>0} $ is suppressed by $e^{-2 |p_0^u| n \pi + \frac{1}{ a^2} n^2 \pi^2 }$, where $n < a^2 |p_0^u| /\pi$.
Similarly, the integral over \( v \) can be evaluated.
Then, because $p_u p_v <0$, the $u$ and $v$ integrations gives the small factor $e^{-\frac{\pi}{2} (|\lambda_0| -\omega_0)}$.
This indeed cancels the huge factor in $  |e^{-i u_0 p^u_0 - i v_0 p^v_0}|=e^{\frac{\pi}{2} (|\lambda_0| -\omega_0)}$.
Therefore, the two-point function can be reproduced even though the normalization factor of the CFT operator is huge.\footnote{
This can also be understood from the Bogoliubov transformation between the AdS-Rindler modes $a_{\omega, \lambda}$ and the global modes $a_{n,m}$. In fact, as shown in Eq.~(3.22) of~\cite{Sugishita:2022ldv}, assuming $\omega < |\lambda|$ and $\lambda > 0$, the saddle point equation
\(
0 = \omega - \lambda - \frac{\omega_{n,m} - n}{\cosh v}
\)
has solutions of the form $v = r + \pi i (1 + 2M)$, where $M$ is an integer and $r$ is an appropriate real constant.
Substituting this into Eq.~(3.21) of~\cite{Sugishita:2022ldv}, namely the factor $e^{i \omega t_R - i \lambda \chi}$, yields an overall factor $e^{- \frac{\pi}{2} (\omega - \lambda)}$, which precisely cancels the exponential factor in $1/N_{\omega,\lambda}$. Note that the shift $v \to v + \pi i$ corresponds to the transformation $\tan \tau \to 1/\tan \theta$, and therefore does not introduce any additional imaginary part.
}

\subsection*{Three-point function in CFT}

We have seen why the two-point function is correctly reproduced, even though it involves a huge normalization factor of $1/N_{\omega, \lambda}$. Next, we investigate the behavior of the three-point function. In particular, we consider configurations in which the operator we are currently studying appears in the middle, as in the following ordering:
\begin{align}
    \langle 0| {\cal O}(u_1,v_1) \, \phi_l^R  \, {\cal O}(u_2,v_2) |0 \rangle.
\end{align}
The three-point function is fixed by the conformal symmetry, except for the overall constant, and given as
\begin{align}
    &\bra{0}  {\cal O}( u_1, v_1) 
     {\cal O}( u, v) 
    {\cal O} ( u_2, v_2) \ket{0} \nonumber \\
    = & 
      C_{{\cal OOO}}
      \left[
64\,
\sinh\frac{u-u_1}{2}\,
\sinh\frac{u-u_2}{2}\,
\sinh\frac{u_1-u_2}{2} \,
\sinh\frac{v-v_1}{2}\,
\sinh\frac{v-v_2}{2}\,
\sinh\frac{v_1-v_2}{2}
\right]^{-\frac{\Delta}{2}},
    \label{3pt}
\end{align}
where $C_{{\cal OOO}}$ is an ${\cal O}(1/N)$ constant.\footnote{
Here, we assume  $C_{{\cal OOO}}$ is not zero. Instead of the three-point function with the scalar primaries, we can consider the one with the stress tensor and the scalar primary.
For this case, we need to use
\begin{align}
    \bra{0}  T ( u_1)  {\cal O}( u, v) 
    {\cal O} ( u_2, v_2) \ket{0} 
    = 
      \frac{\Delta}{2} 
      \frac{1}{(u_1-u)^2(u_1-u_2)^2}
    \frac{1}{(u-u_2)^{ \Delta-2} (v_2-v)^{ \Delta} }
    ,
\end{align}
in the plane coordinate, which is related to the Rindler coordinate by a conformal map.
It is almost evident that the discussions below can be applied to this case, albeit with some differences in numerical factors.
}
Then, as for the two-point function, we find 
\begin{align}
   & \langle 0| {\cal O}(u_1,v_1) \, \phi_l^R  \, {\cal O}(u_2,v_2) |0 \rangle =
     \frac{a^2 \Gamma (\nu+1)}{4 \pi^3} \left( C_{\xi_0, \omega_0, \lambda_0}+ \frac{\psi_{\xi_0, \omega_0, \lambda_0}^{(1,0)}}{a^2} \frac{\partial}{\partial \omega_0}+\cdots 
   \right) \nonumber \\
   &\times  \int^{\infty}_{-\infty} d u \int^{\infty}_{-\infty} d v \left( e^{i (u-u_0) p^u_0 + i (v-v_0) p^v_0-\frac{1}{4 a^2} ((u-u_0)^2+(v-v_0)^2) } \bra{0}  {\cal O}( u_1, v_1) 
     {\cal O}( u, v) 
    {\cal O} ( u_2, v_2) \ket{0} 
   \right).
\end{align}
Because of the $i\epsilon$-prescription for the ordering of the operators, the nearest singularities in the fundamental strip are at
$ u=u_1-i\epsilon$, $ u=u_2+i\epsilon$, $ v=v_1-i\epsilon$, and $ v=v_2+i\epsilon$ with $\epsilon >0$.
However, since the correlator is written in terms of the Rindler functions $\sinh((u-u_i)/2)$ and $\sinh((v-v_i)/2)$, these singularities are accompanied by their images,
\begin{align}
    u&=u_1-i\epsilon+2\pi i n, &
    u&=u_2+i\epsilon+2\pi i n, \nonumber \\
    v&=v_1-i\epsilon+2\pi i n, &
    v&=v_2+i\epsilon+2\pi i n,
    \qquad n\in {\bf Z} .
    \label{sing-lattice}
\end{align}
For non-integer $\Delta/2$, these singularities should be understood as branch points rather than poles, as in the two-point function.
We will shift the integration contour for \(u\) to \(u \in \mathbb{R} + 2 i  a^2 p_0^u\) and for \(v\) to \(v \in \mathbb{R} + 2 i  a^2 p_0^v\), then the contributions from the singularities are dominant.
For the case $p_0^u >0$ and $p_0^v <0$, the nearest singularities at $ u=u_2+i\epsilon$ and $v=v_1- i\epsilon$ are selected for the $u$ and $v$ integrations, respectively.
The remaining crossed singularities are their images and are exponentially smaller.  For example, the image
$u=u_2+i\epsilon+2\pi i n$ with $n\geq1$ gives, relative to the nearest one, a factor of the form
\begin{align}
   \exp\left[-2\pi n |p_0^u|+{\pi^2 n^2\over a^2}+O(1)\right]
   \label{image-supp-u}
\end{align}
for the images crossed by the contour, namely for $2\pi n \lesssim 2a^2|p_0^u|$.
The exponent in \eqref{image-supp-u} is negative in this range, and the endpoint region is suppressed by the same Gaussian factor that suppresses the shifted contour.  The images associated with $u=u_1-i\epsilon+2\pi i n$ and the corresponding images in the $v$ plane are treated in the same way.  Thus the image singularities do not modify, and in particular cannot cancel, the leading exponential dependence obtained from the nearest singularities.
Therefore, keeping the nearest singularities, we have
\begin{align}
   & \langle 0| {\cal O}(u_1,v_1) \, \phi_l^R  \, {\cal O}(u_2,v_2) |0 \rangle =  
    \frac{a^2 \Gamma (\nu+1)}{4 \pi^3} 
    \left(\sinh \frac{u_1-u_2}{2} \sinh\frac{v_2-v_1}{2}  
    \right)^{ -\frac{\Delta}{2}} \nonumber \\
   &\times  
    C_{{\cal OOO}} \left( \frac{2 \pi i}{\Gamma(\frac{\Delta}{2})} \right)^2 \left( C_{\xi_0, \omega_0, \lambda_0}+ \frac{\psi_{\xi_0, \omega_0, \lambda_0}^{(1,0)}}{a^2} \frac{\partial}{\partial \omega_0}+\cdots 
   \right) \nonumber \\
   &\times   
     \left. \frac{\partial^{\frac{\Delta}{2}-1}}{\partial u^{\frac{\Delta}{2}-1}}
    \left( e^{i (u-u_0) p^u_0 -\frac{1}{4 a^2} (u-u_0)^2 } \left(
      \frac{1}{u_1-u }  
    \right)^{ \frac{\Delta}{2}}
    \right) \right|_{u=u_2} \nonumber \\
   &\times \left. \frac{\partial^{\frac{\Delta}{2}-1}}{\partial v^{\frac{\Delta}{2}-1}}
   \left( e^{ i (v-v_0) p^v_0-\frac{1}{4 a^2} (v-v_0)^2 }    
     \left(
     \frac{1}{v-v_2 } 
    \right)^{ \frac{\Delta}{2}}
   \right) \right|_{v=v_1} 
   ,
   \label{r1}
\end{align}
where we have assumed $\Delta/2$ is an integer. Even in the case without this assumption, we can proceed following \cite{Terashima:2023mcr}, and the result will not change.
Then, from this expression \eqref{r1}, we see that if the coordinates \( (u_i, v_i) \) are far from any of the configurations satisfying \( u_2 = u_0 \), \( v_1 = v_0 \), \( u_2 = u_1 \), or \( v_1 = v_2 \),\footnote{
The excluded configurations are those in which one of the insertions is null-related to $(u_0,v_0)$, or the two insertions are null-related to each other. Away from these configurations, the condition is generic.
} 
then the magnitude of the three-point function is essentially determined by the factor \( C_{{\cal OOO}} \, e^{-i u_0 p_0^u-i v_0 p_0^v} \sim \frac{1}{N} \, e^{\frac{\pi}{2} (|\lambda_0|-\omega_0)}\), i.e.
\begin{align}
   & \langle 0| {\cal O}(u_1,v_1) \, \phi_l^R  \, {\cal O}(u_2,v_2) |0 \rangle \sim  \frac{1}{N} \, e^{\frac{\pi}{2} (|\lambda_0|-\omega_0)},
   \label{lc}
\end{align}
where we have omitted the factors which do not affect the leading exponential dependence on $|\lambda_0|-\omega_0$.  These factors include
$C_{\xi_0,\omega_0, \lambda_0} \sim 1/(\omega_0)^{\nu+1/2}$, powers of $p_0^u$ and $p_0^v$ coming from the residues or branch-cut integrals, and finite functions of the insertion points.  They do not contain an additional factor of the form $e^{c(|\lambda_0|-\omega_0)}$ with nonzero constant $c$.  The additional image singularities in \eqref{sing-lattice} are also smaller than the nearest-singularity contribution by factors such as \eqref{image-supp-u}, and hence do not change this leading exponential behavior.
This implies that, for $|\lambda_0|-\omega_0 = \frac{2}{ \pi} c_0 \ln N $,
we have $\langle 0| {\cal O}(u_1,v_1) \, \phi_l^R  \, {\cal O}(u_2,v_2) |0 \rangle \sim  N^{c_0-1}$, and
for $|\lambda_0|-\omega_0 \sim (\ln N)^2 $,
we have $\langle 0| {\cal O}(u_1,v_1) \, \phi_l^R  \, {\cal O}(u_2,v_2) |0 \rangle \sim  N^{c_0 \ln N-1}$, where $c_0$ is a positive constant.
Therefore, the UV cutoff of the bulk theory in the AdS-Rindler patch may have to be $\ln N$, instead of the Planck mass $M_{pl}\sim N^{2/(d-1)}$.

\subsection*{Corrections to the reconstructed operator}

The reconstructed operator \(\phi_l^R\) above is the \(N=\infty\) result, and in the full theory, one may try to improve it by adding \(1/N\) corrections:
\begin{equation}
\phi_l^R \;\to\; \phi_l^R+\delta \phi_l^R .
\end{equation}
One may then ask whether such corrections can remove the exponentially enhanced
contribution in the three-point function. Below, we explain why this does not happen
within the standard perturbative \(1/N\) expansion.

First, let us consider the single-trace sector. At \(N=\infty\), the AdS-Rindler reconstruction
for a given bulk mode is fixed by the Fourier transform \eqref{a_by_O}; in this sense, the
single-trace reconstruction kernel is unique. Therefore, a putative single-trace correction
can only amount to a \(1/N\)-suppressed redefinition of the smearing kernel. Such a correction
changes the coefficient of the reconstructed operator, but does not alter the fact that the
relevant mode carries the large factor \(1/N_{\omega,\lambda}\sim e^{\frac{\pi}{2}(|\lambda|-\omega)}\).
Hence, a perturbative single-trace correction cannot remove the leading exponential enhancement
found in \eqref{lc}.

Next, let us consider the multi-trace sector, as in the usual HKLL-type \(1/N\) corrections.
For example, if \(\delta \phi_l^R\) contains a product of two single-trace operators, then
\begin{equation}
\langle 0|\, O(u_1,v_1)\, \delta \phi_l^R\, O(u_2,v_2)\, |0\rangle
\end{equation}
is determined by a four-point function. Its analytic structure and coordinate dependence are therefore different from those of the three-point function in \eqref{3pt}. In particular, in order
to cancel the term in \eqref{lc}, the correction would have to reproduce the same singularity
structure and the same dependence on \((u_i,v_i)\) as the three-point function for generic
separated insertions. A generic multi-trace contribution does not have this form. Even if two
insertions inside \(\delta \phi_l^R\) are brought close to each other, the resulting OPE merely
produces an effective local single-trace contribution, and therefore reduces to the previous case
of a \(1/N\)-suppressed kernel redefinition.

Thus, within the ordinary perturbative \(1/N\) expansion built from single-trace and finitely
many multi-trace corrections, the exponentially enhanced contribution in \eqref{lc} cannot
be canceled. This is the sense in which the reconstructed operator ceases to be well-defined
once \(|\lambda|-\omega\) becomes of order \(\ln N\).

\section*{Acknowledgement}

The author would like to thank S. Sugishita for collaboration in the early stages of this work
and helpful discussions and comments.
The author thank T. Kawamoto, N. Tanahashi, and S. Yoshikawa for the useful comments.
This work was supported by MEXT-JSPS Grant-in-Aid for Transformative Research Areas (A) ``Extreme Universe'', No. 21H05184.
This work was supported by JSPS KAKENHI Grant Number 	24K07048.

\hspace{1cm}


\appendix

\section{Asymptotic behavior }
\label{ab}

In this Appendix, we will evaluate the radial wavefunction $\tilde{\psi}_{\omega, \lambda}(\xi)$, given in \eqref{psiR}, for large $\omega,|\lambda|$.
This requires analyzing the asymptotic behavior of the Gauss hypergeometric function, as has been done in \cite{paris2013asymptotics} \cite{Cvitkovic2017asymptotic}. However, these analyses are not applicable to the parameter region of interest here; therefore, in the following, we use a slightly different integral representation.

The Gauss hypergeometric function is defined by the series as
\[
{}_2F_1(a, b; c; x) = \sum_{n=0}^{\infty} \frac{(a)_n \, (b)_n}{(c)_n} \cdot \frac{x^n}{n!},
\]
where $(q)_n$ is the Pochhammer symbol:
$
(q)_n = \frac{\Gamma(q+n)}{\Gamma(q)}.
$
We will consider the asymptotic behavior of 
\[ {}_2F_1(a+ i \epsilon \tilde{\lambda}, b-i \tilde{\lambda}; c; x) \]
for large $ \tilde{\lambda}$, 
where
\begin{align}
    x=\frac{1}{1+ \xi^2}, \,\,\, a=b=\frac{c}{2}=\frac{\nu+1}{2},  \,\,\,\, \tilde{\lambda} = -\frac{1}{2} (\omega+\lambda), \,\,\epsilon=\frac{\lambda-\omega}{\lambda+\omega},
\end{align}
which are real, in particular $0 < x \leq 1$.

By the Euler integral expression, which is valid for $c>b>0$ and $0 \leq x \leq 1$, we find
\begin{align} 
   {}_2F_1(a+ i \epsilon \tilde{\lambda}, b-i \tilde{\lambda}; c; x)
   &= \frac{\Gamma(2b)}{\Gamma(b-i \tilde{\lambda}) \Gamma(b+i \tilde{\lambda})} \int_0^1 dt \,\, t^{b-i \tilde{\lambda}-1} (1-t)^{b+i \tilde{\lambda}-1} (1-t x)^{-b-\epsilon i \tilde{\lambda}} \nonumber \\
   &= \frac{\Gamma(2b)}{\Gamma(b-i \tilde{\lambda}) \Gamma(b+i \tilde{\lambda})} \int_0^1 dt f(t) e^{i \tilde{\lambda} \psi(t)},
   \label{int}
\end{align}
where
\begin{align}
    f(t)=\left( \frac{t}{1-t} \right)^{b-1} (1-t x)^{-b}, \,\,\,\,\, \psi(t) =-\ln t +\ln (1-t) -\epsilon \ln (1-tx),
\end{align}
and $(1-t x)^{-b-\epsilon i \tilde{\lambda}}$ is defined such that  $\lim_{t \rightarrow 0} (1-t x)^{-b-\epsilon i \tilde{\lambda}}=1$.
We take the branch cut along \( t \in (-\infty, 0] \) and \( t \in [1, 1/x] \).
The derivative of the phase is given by
\begin{align}
    \frac{d}{d t}\psi(t) =-\frac{1}{t(1-t)} + \epsilon x \frac{1}{1-tx} =\frac{tx-1+\ep xt(1-t)}{t(1-t)(1-tx)}, 
    \label{dp}
\end{align}
which vanishes at the following two points:
\begin{align}
    t_{\pm} = \frac{1}{2} \left( 1+\frac{1}{\epsilon} \right) \left( 1 \pm \sqrt{1-\frac{4 \epsilon}{(1+\epsilon)^2 x}}\right),
\end{align}
and we easily find 
\begin{align}
    \left. \frac{d^2}{d^2 t}\psi(t) \right|_{t=t_-} =- \epsilon x \frac{(t_- -t_+)}{t_-(1-t_-)(1-t_- x)}, 
\end{align}
and $\left. \frac{d^2}{d^2 t}\psi(t) \right|_{t=t_+} = \epsilon x \frac{(t_- -t_+)}{t_+(1-t_+)(1-t_+ x)}$. 

If the saddle point approximation for large $|\tilde{\lambda}|$ is valid and both of the two saddle points contribute,
we have
\begin{align}
   {}_2F_1(a, b; c; x) & \simeq \frac{\Gamma(2b)}{\Gamma(b-i \tilde{\lambda}) \Gamma(b+i \tilde{\lambda})} 
   \left(  f(t_-) e^{i \tilde{\lambda} \psi(t_-) - i \theta^{(t_-)}} \sqrt{\frac{2 \pi}{ \left|\tilde{\lambda} \left. \frac{d^2}{d^2 t}\psi(t) \right|_{t=t_-} \right| }} +(t_- \rightarrow t_+) \right)
   \nonumber \\
   &\simeq \frac{\Gamma(2b)}{2 \pi i }  \frac{  e^{\pi |\tilde{\lambda}| 
   } }{
   |\tilde{\lambda}|^{2b-1}}
 \left(  f(t_-) e^{i \tilde{\lambda} \psi(t_-) - i \theta^{(t_-)} } \sqrt{\frac{2 \pi}{\left|\tilde{\lambda} \left. \frac{d^2}{d^2 t}\psi(t) \right|_{t=t_-} \right| }} +(t_- \rightarrow t_+) \right),
 \label{asy}
\end{align}
where
\[
\theta^{(t_-)}
 \;=\;
\frac{\arg \left( \tilde{\lambda}\left. \frac{d^2}{d^2 t}\psi(t) \right|_{t=t_-}  \right)}{2}
\;+\;
\frac{3 \pi}{4}
\pmod{\pi}
\]
and the mod $\pi$ ambiguity is determined by the direction of the descent path near the saddle. 
Which saddle point actually contributes depends on the values of the parameters. In the following, we analyze each case separately, in particular focusing on the large $\tilde{\lambda}$ behavior.

First, we will consider the case with 
\begin{align}
    |\lambda| - \omega > 0,
\end{align}
 which implies $\epsilon >0$.
We also assume $\lambda <0$, which means $\epsilon >1$.\footnote{
We can evaluate it for $\lambda >0$, which means $\epsilon <1 $, by using the identity \(
{}_2F_1(a, b; c; x)=
{}_2F_1(b, a; c; x)
\), which gives the transformation $\lambda \rightarrow -\lambda$.  
}

We also take 
\begin{align}
    x > x_* \equiv \frac{4 \epsilon}{(1+\epsilon)^2}.
    \label{xx}
\end{align}
This case corresponds to the region that is relevant for the wave packet under consideration
because the condition \eqref{xx} is equivalent to  \eqref{hhc}.
Then, the two solutions are real and satisfy
\begin{align}
    0 < \frac{1}{\epsilon} < t_- < \frac{1}{2} \left( 1+\frac{1}{\epsilon} \right) < t_+ <1.
\end{align}
In particular, we see that $t_{\pm} \rightarrow \frac{1}{2} \left( 1+\frac{1}{\epsilon} \right)$ for $x \rightarrow x_*$
and $t_+ \rightarrow 1$, $t_- \rightarrow 1/\epsilon$ for $x \rightarrow 1$. 
We also see $\psi(t_\pm)$ is real, which implies both saddles potentially contribute dominantly.
Indeed, we will see below that both saddles contribute
and the factor $e^{i \tilde{\lambda} \psi(t_\pm)}$ in \eqref{asy} is just a phase.
Thus, the large $\tilde{\lambda}$ behavior of $ |\tilde{\psi}_{\omega, \lambda}(\xi)| $ in \eqref{psiR} is
\begin{align}
    |\tilde{\psi}_{\omega, \lambda}(\xi)| \sim 
    e^{-\frac{\pi}{2} (|\lambda|-\omega)} (\lambda^2-\omega^2)^{\nu/2} e^{\frac{\pi}{2} (|\lambda|-\omega)} (|\lambda|-\omega)^{-(2b-1+1/2)} 
    =\left( \frac{|\lambda|+\omega}{|\lambda|-\omega}\right)^{\frac{\nu}{2}} 
    \frac{1}{(|\lambda|-\omega)^{\frac{1}{2}}}
    ,
    \label{psitab}
\end{align}
where we neglected the $\epsilon$ dependence.
It is important to note that the exponential factors canceled each other for $ |\tilde{\psi}_{\omega, \lambda}(\xi)| $.
More precisely, the asymptotic expansion is given by 
\begin{align}
\label{psiN}
{\widetilde\psi_{\omega,\lambda}(\xi)\over N_{\omega,\lambda}}
\simeq&\,
{\Gamma(2b)\over \sqrt{2\pi}\,i\,\Gamma(\nu+1)}
(1+\xi^2)^{-\Delta/2}
\nonumber\\
&\times
\left[
{ e^{{\pi\over2}(|\lambda|-\omega)(1+i\psi(t_-))
      +i\omega\ln\left({\xi\over\sqrt{1+\xi^2}}\right)}
 \over ((|\lambda|-\omega)/2)^{\nu+1/2} }
{ f(t_-)e^{-i\theta^{(t_-)}}
 \over \sqrt{\left|\left.{d^2\psi\over dt^2}\right|_{t=t_-}\right|} }
+(t_-\to t_+)
\right] .
\end{align}

For 
\begin{align}
    x < x_* \equiv \frac{4 \epsilon}{(1+\epsilon)^2},
    \label{xx2}
\end{align}
the two solutions are not real and $t_-=(t_+)^*$,
which implies $\psi(t_-)=(\psi(t_+))^*$.
Thus, one of $e^{i \tilde{\lambda} \psi(t_-)}$ and $e^{i \tilde{\lambda} \psi(t_+)}$ becomes exponentially large, while the other becomes exponentially small, and then only the exponentially small term in \eqref{asy} can contribute
for this case because the integral in \eqref{int} cannot give an exponentially large factor.
This means that the large $\tilde{\lambda}$ behavior of $ |\tilde{\psi}_{\omega, \lambda}(\xi)| $ in \eqref{psiR} is
exponentially small:
\begin{align}
    |\tilde{\psi}_{\omega, \lambda}(\xi)| \sim 
    |\lambda|^{-\frac12} e^{-|\tilde{\lambda} \,\,\operatorname{Im} (\psi(t_-))|},
\end{align}
as expected.

Next, we will consider the case with 
\begin{align}
    |\lambda| - \omega < 0,
\end{align}
 which implies $\epsilon <0$.
We also assume $\lambda > 0$, which means $-1 <\epsilon <0 $.\footnote{
We can evaluate it for $\lambda <0$, which means $\epsilon <-1$, by using the identity \(
{}_2F_1(a, b; c; x)=
{}_2F_1(b, a; c; x)
\), which gives the transformation $\lambda \rightarrow -\lambda$.  
}
For this case, the two solutions are real and satisfy
\begin{align}
   t_+ < 0, \,\,\,  1 <t_- < \frac{1}{x},
\end{align}
because $t_-|_{x=1}=1$ and $\frac{\partial}{\partial x}(t_--1/x)=\frac{1}{x^2}(1+\frac{1}{1+\epsilon} \frac{1}{\sqrt{1-\frac{4 \epsilon}{(1+\epsilon^2) x}}})>0$.
Then, the imaginary part of the phase is 
given as $\operatorname{Im}(\psi(t_+)) =-\pi$ and $\operatorname{Im}(\psi(t_-)) =- \pi$, where we choose the branch of the $\ln t$ or $\ln(1-t)$ such that the integral gives an exponentially small and maximal and both saddles potentially contribute dominantly.
Indeed, we will see below that both saddles contribute
and the factor $|e^{i \tilde{\lambda} \psi(t_\pm)}| =e^{-\pi |\tilde{\lambda}|}$ in \eqref{asy} is just a phase.
Thus, the large $\tilde{\lambda}$ behavior of $ |\tilde{\psi}_{\omega, \lambda}(\xi)| $ in \eqref{psiR} is
\begin{align}
    |\tilde{\psi}_{\omega, \lambda}(\xi)| \sim 
     |\lambda|^{-\frac12}. 
\end{align}


\subsection{Explicit descent path}

Below, we will consider which saddles indeed contribute.

First, we will consider the case with 
\begin{align}
    |\lambda| - \omega > 0, x> x_*,\, \lambda <0.
    \label{c1}
\end{align}
The two saddles for this case are on the original path of the integral.
We can deform the path slightly such that $t=t_- + e^{i\pi/4} s$ near $t_-$ and $t=t_+ + e^{-i\pi/4} s$ near $t_+$,
where $s$ is a real parameter. 
The entire path is chosen appropriately so that it becomes a smooth, small deformation of the original path connecting 0 and 1.
Then, we find that $e^{i\tilde{\lambda} \psi(t)}$ takes its maximal value at the saddle points, since for small $\delta$ we have
\[
\psi(t + i \delta) \simeq \psi(t) + i \delta \, \frac{(-\epsilon x)(t - t_-)(t - t_+)}{t(1 - t)(1 - t x)},
\]
and the factor $\frac{(-\epsilon x)(t - t_-)(t - t_+)}{t(1 - t)(1 - t x)}$ has a definite sign in each of the intervals $t \in (0, t_-)$, $t \in (t_-, t_+)$, and $t \in (t_+, 1)$.
Therefore, the two saddles contribute dominantly to the integrals.

To determine which saddle points contribute dominantly, we can also use the Lefschetz thimble method, where we analyze the structure of the complex flow associated with the action.
We define the action as \( S(t) = i \psi(t) \). 
To investigate the structure of the Lefschetz thimbles, we study the complex flow generated by the vector field \( \psi'(t) \). 
The flow lines (streamlines) correspond to the integral curves of the gradient descent equation
\[
\frac{dt}{d\tau} = \overline{\left( \frac{dS}{dt} \right)} = i \overline{ \psi'(t) },
\]
which describe the steepest descent paths in the complex \( t \)-plane.

To visualize the behavior of the flow and identify the contributing saddle points, 
we will plot the streamlines of \( \psi'(t) \) for representative values of the parameters.
This plot reveals the thimble structure and helps determine which saddles give dominant contributions to the integral.
In particular, whether a given saddle point contributes is determined by the intersection number 
between the original integration contour and the dual thimble (the upward flow from the saddle). 
We read off these intersection numbers directly from the streamline plots, 
which clearly show how the original contour decomposes into a sum over thimbles.

\begin{figure}[htbp]
  \centering
  \includegraphics[width=0.5\textwidth]{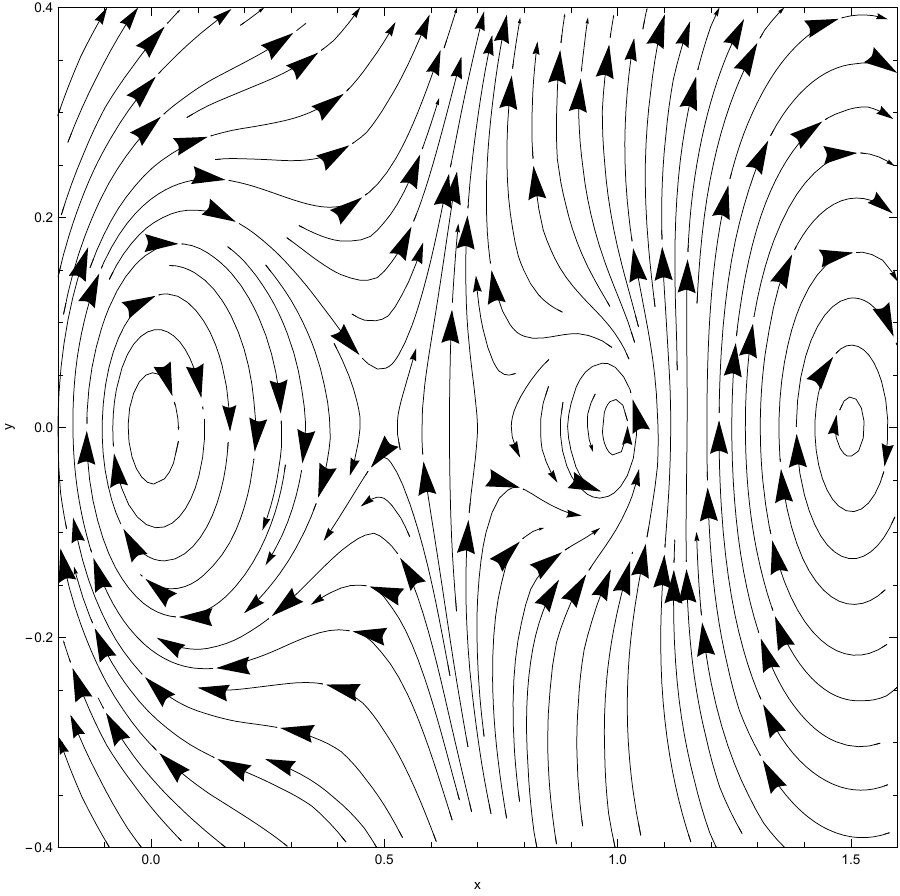}
  \caption{
    Streamline plot of the complex vector field defined by the derivative
    \(
      i \psi'(t) = i\left(-\frac{1}{t(1 - t)} + \frac{\epsilon x}{1 - x t} \right)
    \)
    with parameters \(\epsilon = 4\) and \(x = \tfrac{2}{3}\), evaluated at \(t = {\rm x} - i {\rm y}\).
    The vector field is given by \((\operatorname{Re}[i \psi'(t)], \operatorname{Im}[i \psi'(t)])\), and plotted on the domain \({\rm x} \in [-0.2, 1.6]\), \({\rm y} \in [-0.4, 0.4]\).
    Notable singularities arise at the branch points \(t = 0\), \(t = 1\), and \(t = \tfrac{3}{2}=\frac{1}{x}\), around which the flow lines exhibit characteristic structures.
  }
  \label{fig:vectorFieldPlot}
\end{figure}

In Fig. \ref{fig:vectorFieldPlot}, we see such a streamline plot for a case with $ |\lambda| - \omega > 0, \, x > x_*$.
For this case, we see that both of the saddles contribute dominantly because the upward flow from two saddles intersects with the original integration contour, ${\rm x} \in [0,1], \ {\rm y}=0$.

\begin{figure}[htbp]
  \centering
  \includegraphics[width=0.5\textwidth]{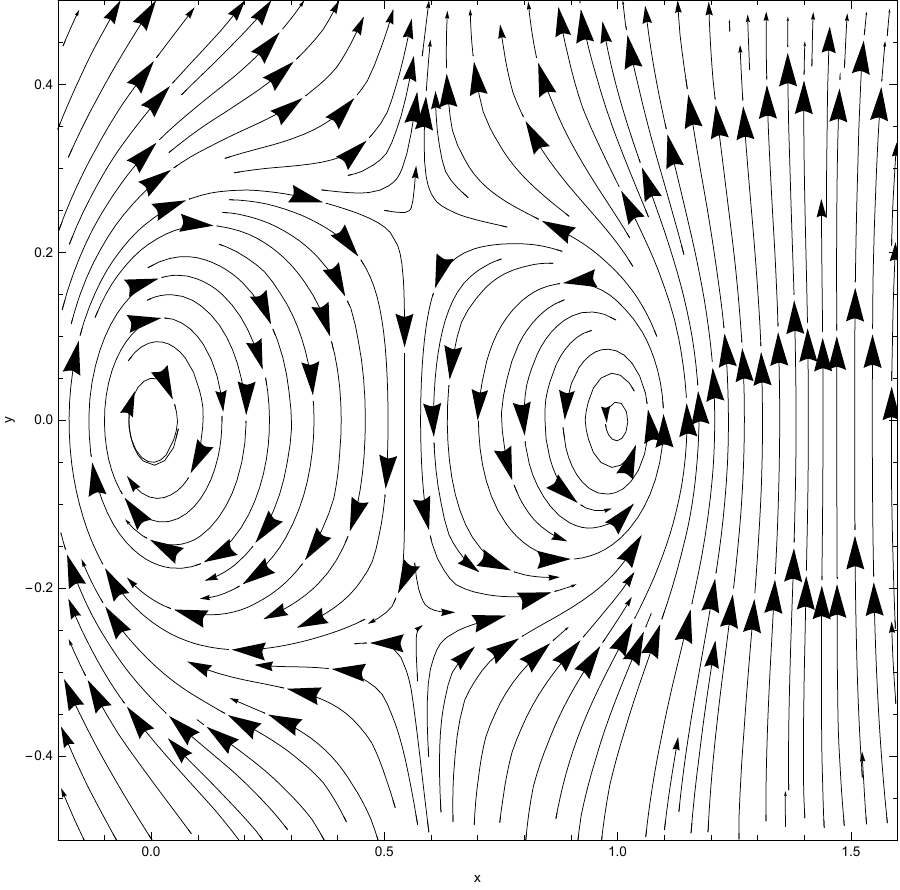}
  \caption{
    Streamline plot 
    shown over the domain \({\rm x} \in [-0.2, 1.6]\), \( {\rm y} \in [-0.5, 0.5]\) with parameters \(\epsilon = 8\), \(x = \tfrac{1}{3}\).
  }
  \label{fig:streamplot-epsilon8-w13}
\end{figure}

In Fig. \ref{fig:streamplot-epsilon8-w13}, we see such a streamline plot for a case with $ |\lambda| - \omega > 0, x< x_*$.
For this case, we see that only one saddle contributes dominantly because the upward flow from one saddle does not intersect with the original integration contour, ${\rm x} \in [0,1], \, {\rm y}=0$.

\begin{figure}[htbp]
  \centering
  \includegraphics[width=0.5\textwidth]{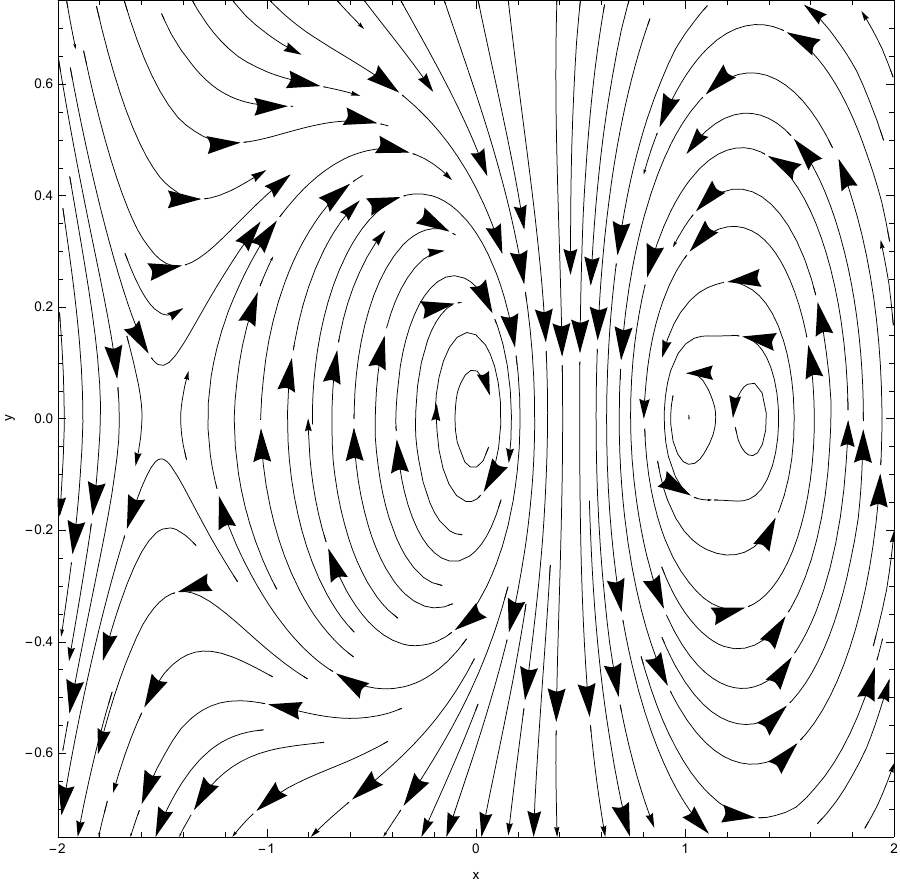}
  \caption{
    Streamline plot shown over the domain
    \( {\rm x} \in [-2.0, 2.0]\), \({\rm y} \in [-0.75, 0.75]\), with parameters \(\epsilon = -\tfrac{3}{4}\), \(x = \tfrac{3}{4}\).
  }
  \label{fig:streamplot-em075-w075}
\end{figure}

In Fig. \ref{fig:streamplot-em075-w075}, we see such a streamline plot for a case with $ |\lambda| - \omega < 0$.
For this case, we see that both of the saddles contribute dominantly because the upward flow from two saddles intersects with the original integration contour, ${\rm x} \in [0,1], \ {\rm y}=0$.

\paragraph*{Locally flat approximation of the wave function}
Here, we will solve the wave equation 
$(\square -m^2) v_{\omega,\lambda}=0$, 
where $v_{\omega, \lambda}=e^{- i\omega t_R+ i \lambda \chi} \, \tilde{\psi}_{\omega, \lambda}(\xi)$
near the point $t_R=t_R^0, \chi=\chi_0, \xi=\xi_0$
using the locally flat coordinate:
$\hat{t}=\xi_0 (t_R-t_R^0), \hat{\xi}=\frac{\xi-\xi_0}{\sqrt{1+(\xi_0)^2}}, \hat{\chi}=\sqrt{1+(\xi_0)^2} (\chi-\chi_0)$.
Then, we easily find $v_{\omega, \lambda}$ is approximately given by a linear combination of $e^{- i \hat\omega \hat{t}+ i \hat\lambda \hat\chi \pm i \sqrt{\hat\omega^2-\hat\lambda^2} \hat\xi}$, where
$\hat\omega =\frac{\omega}{\xi_0}, \hat\lambda=\frac{\lambda}{\sqrt{1+(\xi_0)^2}}$.
Here, if $1 >\frac{\hat\omega^2}{\hat\lambda^2}=\frac{1+(\xi_0)^2}{(\xi_0)^2} \frac{\omega^2}{\lambda^2}$, the wave function is not wave-like, but exponential damping or increasing and we expect that  
$e^{- i \hat\omega \hat{t}+ i \hat\lambda \hat\chi -\sqrt{\hat\lambda^2-\hat\omega^2} \hat\xi} $ should be chosen because it will decay for $\xi \rightarrow \infty$. As expected, this condition is the same as \eqref{hhc}.
For the wave-like case, because the mode should satisfy  
\begin{align}
    \int^{\infty}_0 \!d \omega\, 2\omega\,  \tilde{\psi}_{\omega, \lambda}(\xi) 
     \tilde{\psi}_{\omega, \lambda}(\xi') 
     =\frac{\xi}{(1+\xi^2)^\frac{d-2}{2}}\delta(\xi-\xi'), 
     \label{norm_xi}
\end{align}
the coefficients of the linear combination of $e^{- i \hat\omega \hat{t}+ i \hat\lambda \hat\chi \pm i \sqrt{\hat\omega^2-\hat\lambda^2} \hat\xi}$
will be ${\cal O}(\omega^{-\frac{1}{2}})$ for large $\omega$ with a fixed $\lambda$.
These are consistent with the previous analysis of the asymptotic behavior of the Gauss hypergeometric function.


\newpage 

\bibliographystyle{utphys}
\bibliography{main.bib}
\end{document}